\documentclass[twocolumn,english,prb,superscriptaddress]{revtex4}
\usepackage[T1]{fontenc}
\usepackage[latin1]{inputenc}
\usepackage{graphicx}

\makeatletter

\usepackage{graphicx}

\usepackage{graphicx}

\usepackage{babel}
\makeatother
\begin{document}

\title{Nanotechnology and Society: A discussion-based undergraduate course}

\author{Charles Tahan}

\email{charlie@tahan.com}

\affiliation{Department of Physics, University of Wisconsin, Madison, Wisconsin
53706}

\author{Ricky Leung}

\affiliation{Department of Sociology, University of Wisconsin, Madison, Wisconsin
53706}

\author{G. M. Zenner}

\affiliation{Materials Research Science and Engineering Center, University of
Wisconsin, Madison, Wisconsin 53706}

\author{K. D. Ellison}

\affiliation{Graduate School, University of Wisconsin, Madison, Wisconsin 53706}

\author{W. C. Crone}

\affiliation{Department of Engineering Physics, University of Wisconsin, Madison,
Wisconsin 53706}

\author{Clark A. Miller}

\affiliation{LaFollette School of Public Affairs, University of Wisconsin, Madison,
Wisconsin 53706}

\begin{abstract}
Nanotechnology has emerged as a broad, exciting, yet ill-defined field
of scientific research and technological innovation. There are important
questions about the technology's potential economic, social, and environmental
implications. We discuss an undergraduate course on nanoscience and
nanotechnology for students from a wide range of disciplines, including
the natural and social sciences, the humanities, and engineering.
The course explores these questions and the broader place of technology
in contemporary societies. The course is built around active learning
methods and seeks to develop the students' critical thinking skills,
written and verbal communication abilities, and general knowledge
of nanoscience and nanoengineering concepts. Continuous assessment
was used to gain information about the effectiveness of class discussions
and enhancement of student understanding of the interaction between
nanotechnology and society. 
\end{abstract}
\maketitle

\section{Introduction}

Nanotechnology is cool. This truth has great allure to students and
educators both. As public attention to nanoscale science and engineering
spotlights research and the potential of new discoveries, students
are pulled toward careers in science, engineering, and related social
sciences or businesses. Educators not only have a new field of endeavor
and questions to explore, but also another hook to gain the attention
and interest of students. Nanoscale science and engineering raises
many important questions, especially at the intersection of technology
and society. Government funding of the field, which includes funds
specifically earmarked for environmental and societal impact studies,\cite{Roco-Broad,Roco-NNI}
shows that policy officials are focussed on addressing these societal
concerns. The ability to create nanoscale materials and devices will
generate new ways for people to understand and exploit nature. But
who will have access to these new capabilities? How will they be applied?
By whom? What are the consequences for our society?

It is incumbent on science and engineering educators to partner with
their counterparts in the social sciences and public policy to bring
the discussion about the connections between technology and society
to undergraduate students. Before this course, a curricular gap existed
in nanoscale science and engineering education at the University of
Wisconsin-Madison (UW). Nanotechnology education has primarily focused
on the field's technical aspects, with little emphasis on issues such
as the social and ethical implications of design choices, public attitudes
toward new technologies, and nanotechnology policy.

A course on nanotechnology and its societal implications can serve
multiple purposes. Recruitment, education, introduction to nanoscale
science and engineering, and science and technology studies (STS)
all fall in its scope. STS itself is an umbrella term for a number
of related topics including the sociology of science knowledge, philosophy
of science, and history of science and technology. Here we describe
a nontechnical course for undergraduates that introduces a broad audience
to nanoscale science and engineering and STS. The course is open to
all majors and satisfies a humanities requirement for undergraduates.
Although designated as a 200-level class (freshmen or sophomores),
the course was open to all students. The course is discussion-based,
requires active student involvement, and focuses on readings, group
discussion sessions, role-playing exercises, essay assignments and
exams, and a semester-long research project with a final presentation.

The course, Nanotechnology and Society, was offered in two sections
in the spring of 2005. Two sections of a STS course, Where Science
Meets Society, were designed and led by a graduate student specifically
trained in nanoscale science and engineering and STS in the previous
semester. In prior versions of the latter course STS topics were covered
in a more general context of many technologies, without including
learning of specific science concepts or facts. The course is regularly
taught as a first-year seminar and satisfies either a humanities or
social sciences requirement within the university's core liberal arts
curriculum. It is well known by first-year advisors in the College
of Letters and Science and the College of Engineering and has proven
successful in drawing students from humanities, science, and engineering.
This year, two sections were separated and designated for the new
course on Nanotechnology and Society. This paper discusses the section\cite{URL-Tahan-NanoSocietyCourse}
taught by co-author Tahan, a physics graduate student; the other section
was taught by co-author Leung, a sociology graduate student. Both
courses were based on a similar core curriculum developed in the prior
semester.\cite{PROC-Crone-NanoSoc}

\section{Preparation}

To develop an effective undergraduate course in nanotechnology and
society, we first needed to educate the educators. To this end, a
seminar was created for advanced graduate students in the sciences,
engineering, humanities, and social sciences to explore questions
about the connections between nanotechnology and societal issues and
to reflect on the broader place of technology in modern societies.
The instructors for this seminar (co-authors Zenner, Ellison, Crone,
and Miller) came from backgrounds in engineering, public policy, and
the humanities. In addition, a partnership was initiated through a
National Science Foundation funded Nanotechnology Undergraduate Education
grant between the Materials Research Science and Engineering Center
and the Robert and Jean Holtz Center for Science and Technology Studies,
a newly established center for research and teaching in the history,
sociology, and philosophy of science, technology, and medicine at
UW.

The seminar was offered to graduate students for either one or three
credits. Students who chose the one-credit option were expected to
attend the seminar's first hour, read and discuss the class materials,
and write a one-page response essay each week. This part of the seminar,
attended by ten graduate students and post-doctoral associates in
the Fall 2004 semester, focused on theories and approaches to understanding
the social dimensions of technology applied to the case study of nanotechnology.
More detailed course information is provided in Refs.~\onlinecite{PROC-Crone-NanoSoc}
and \onlinecite{URL-MRSECNanoSocietyCourses}.

The three credit option had an additional emphasis on the development
of teaching skills and the creation of a teaching portfolio. Students
who chose this option attended a second hour of the seminar and developed
an annotated syllabus for an undergraduate seminar in nanotechnology
and society. This portion of the course was designed for future educators
who wished to teach nanotechnology and society topics, either as a
stand-alone course or as part of another course. These students also
led the discussion in the first hour on a rotating basis, giving them
an opportunity to test various active learning techniques such as
think-pair-share, jigsaw (where the class is divided in parts to solve
a problem), town-meeting formats, group discussion, and blackboard
exercises. This second part of the seminar introduced approaches,
materials, and skills for teaching undergraduates how to think critically
about the social aspects of technology. Four graduate students completed
the three credit course, including the two who taught their own courses
in the spring. One of these courses is described here.

\section{Goals and Course Content}

STS 201, Nanotechnology and Society, set broad goals in both its scope
and content. As stated in the syllabus, the objectives of this course
include the following:

\begin{enumerate}
\item Introduce the broad field of nanotechnology and the basic science
and technology. 
\item Consider the societal implications of nanotechnology in the context
of social, scientific, historical, political, environmental, philosophical,
ethical, and cultural ideas from other fields and prior work. 
\item Develop questioning, thinking, idea producing, and communication skills,
both written and verbal. 
\end{enumerate}
Because STS 201 was primarily a humanities course, the focus was on
understanding the implications of technology and its interactions
with society, specifically applied to nanoscale science and engineering.
>From a deeper curriculum perspective, the goals include the following.

\begin{enumerate}
\item Introduce the various social theories of technology, such as technological
determinism and the social construction of technology. 
\item Explore the wider social, historical, and cultural contexts in which
nanoscale science and engineering are embedded. 
\item Examine the technical and social elements of nanotechnological systems. 
\item Provide skills and resources for learning about the technological
infrastructures of modern societies and the potential impacts of developments
in nanotechnology. 
\item Investigate why people sometimes fear new technologies, including
studies of technological utopias and dystopias, accidents, risk, and
concerns about loss of control. 
\end{enumerate}
An obvious question is how much science was included. Students were
required to learn some of the basic science of the nanotechnologies
discussed in class. We illustrate the level by the example of the
nanotechnology of nanocrystals or quantum dots. The students were
expected to learn some primitive semiconductor physics to understand
why nanoscale semiconductor crystals exhibit new properties, such
as changes in color emission at certain size thresholds. The notion
of a bandgap between core (valence) electron levels and free (conduction)
levels was introduced with a discussion of light (photon) excitation.
Students were expected to learn how the energy gap between the electron
levels changes with decreasing size and the reason (quantum confinement
effects). This understanding was then compared and applied to the
application of quantum dots for medical contrast imaging. Lectures
in addition to books for a lay audience, for example, Refs.~\onlinecite{ART-un,ART-NanoShaping,ART-SwissRe,waser-nano},
provided the main teaching materials.%
\begin{table}
\begin{enumerate}
\item Introduction to Nanotechnology and Society (classes 1--3, essay 1).
How is nanotechnology defined? \setlength{\baselineskip}{12pt}
\item Nanoscience/technology (classes 4, 5, 10, 12, 14, 37--44).

\begin{enumerate}
\item Policy reports and reviews. 
\item Topics: New nanoscale effects; quantum vs.\ classical; Nano-manufacturing;
quantum dots and nanoparticles; carbon; medical applications. 
\item Student research projects and presentations. 
\end{enumerate}
\item Nanotech in Culture (classes 6, 8, 9, 22, 24, 46).

\begin{enumerate}
\item What real nanoproducts are on the market now and what's nanohyped? 
\item How does science fiction bring science/technology to the public? See
Refs.~\onlinecite{Flynn,NewBreed,LearningCurve}.
\item How has nano seeped into the media? 
\end{enumerate}
\item Revolutions and the History of Science and Technology (classes 31,
46, essay 3). Is nanotech a new industrial revolution? 
\item Technology and Society (classes 7, 9, 11, 13, 15, 16, 24, 32, 46,
essay 2).

\begin{enumerate}
\item Do technological innovations necessarily contribute to progress? 
\item How does technology affect the way we live? 
\item How do the users shape the development of technology? 
\item Is technology political? 
\end{enumerate}
\item How Government Drives Technology (classes 23, 25, 46, essay 4).

\begin{enumerate}
\item How much money is being invested nanotechnology and science? 
\item What agencies handle nanotech funding? 
\item How does the military's needs shape our world? 
\end{enumerate}
\item Weighing the Risks (classes 33, 34, 35, 36, 46, essay 4).

\begin{enumerate}
\item How does society decide what kinds of risks are acceptable given the
possible consequences of pursuing a certain technology or science? 
\item Is nanoscale science and engineering more dangerous than micro? 
\item What is a normal accident? 
\end{enumerate}
\item Thinking About the Future (classes 30, 45, 47).

\begin{enumerate}
\item What do the minds of today (or at least those who get media attention)
think about nanotech? (See for example, Refs.~\onlinecite{Drexler}
and \onlinecite{Mulhall}.) 
\item More Science Fiction. 
\item Reflections. What have we learned? 
\end{enumerate}
\end{enumerate}
\vspace{-0.5cm}

\caption{Course outline. The course materials can be found online.\cite{URL-Tahan-NanoSocietyCourse}\label{cap:CourseOutline}}
\end{table}

The class outline given in Table~\ref{cap:CourseOutline} is mostly
chronological except that the nanoscience subtopics were distributed
throughout the semester instead of at a single time. We began reading
general introductory articles on nanotechnology such as found in popular
science magazines, think-tank and corporate reports, and then began
looking at the STS topics one-by-one, intermixing STS topics with
nanoscale science and engineering. In the last few weeks the students
reported on their research on a specific topic in nanoscale science
and engineering.

The STS readings were introductory in nature (such as in Refs.~\onlinecite{WinnerBook,winner-congress,cross-szostak,GolemScience,GolemTech,handbook,social-shaping,Smith-Military-Noble,Perrow-Normal,Misunderstood,TechandFuture,FiftyYears})
and assumed an audience not familiar with the more complex analytical
techniques and terms that are used in higher level sociology or history
of science courses. The readings for this section are available online.\cite{URL-Tahan-NanoSocietyCourse}
The overall curriculum consisted of components that introduced a concept
in STS and then used STS as a means to apply or interpret the concept.

\section{Requirements and Output}

Because the course was primarily discussion based, class participation
(including homework) was highly valued and vital to exploring the
issues fully. It counted for 25\% of the grade, including the expectation
that students participate or lead group discussions, present before
the class, and participate in debates, mock hearings, or other cooperative
activities. Reading was assigned for nearly every class, but homework
was occasional and included small writing or research assignments
to be shared with the class. An example was an assignment for which
the students chose from a list of professors at the university doing
nanoscale science and engineering research and reported to the class
on the interests of a particular research group. Another assignment
was to find a nanoscale science and engineering product in the news,
learn about it, and teach what they learned to the class.

To a large extent the course was about connecting disparate questions,
concepts, facts, and ideas, and then raising new questions. Writing
is a vital process in this approach to thinking because it is a formal
way of integrating ideas and communicating. There were four, 2--3
page, double-spaced response or op-ed type essays for each of the
main topics (see Table~\ref{cap:Essays}). The four graded essays
counted for a total of 20\% of the grade.%
\begin{table}
\begin{enumerate}
\item You are interviewing for a job at McKinsey, a prestigious consulting
firm. During your interview you mention that you have experience thinking
about the societal implications of technology, specifically nanotechnology.
The interviewer asks you to go home and write a two to three-page
executive summary defining nanotechnology (which she, a non-scientist,
can understand) and suggesting specific areas where McKinsey may be
able to do in the future. You must really impress her to get the job.
\item Does nanotechnology have politics? Make your case, for or against,
using the articles we have talked about in class (see, for example,
Ref.~\onlinecite{WinnerBook}).
\item Is the field of nanotechnology a revolution or just evolution?
\item Write a brief testimony to be presented to the congressional subcommittee
reviewing the National Nanotechnology Initiatives and address the
following questions. Should the government continue funding of research
in nanotechnolog? In what specific areas? How? Should the public be
brought into the nanotech development process? How? You will represent
a specific political group, for example, the military or AAAS.
\end{enumerate}
\vspace{-0.5cm}

\caption{Essay assignments (abbreviated).\label{cap:Essays}}
\end{table}

Two formal exams counted for another 25\% of the grade. The remaining
30\% of the course requirements was assessed from individual research
projects and class presentations. A list of topics was developed by
the instructor, and each student selected one and become the class
{}``expert'' on it. These topics provided a means to explore in
more depth some of the subfields of nanoscale science and engineering
and allowed the students to teach each other instead of sitting through
lectures by the instructor. The goal was to produce a pamphlet on
key nanotechnologies circa 2005 that may have value to future iterations
of the class and to the public. It also provided an opportunity for
more advanced students to contribute their particular expertise that
might be outside the realm of the instructor's specialty. Approximately
two-thirds of each roughly five double-spaced page report covered
the science of the selected topic with the last one-third on the societal
implications. Each student also gave a 20 minute PowerPoint or blackboard
presentation. Examples of the nanotopics include nano-nuclear batteries,
nanotechnology and cancer, nanofiltration, and nanotechnology and
agriculture. The student reports and presentations are also available.\cite{URL-Tahan-NanoSocietyCourse}

\section{Assessment}

In addition to the traditional evaluation of student work discussed
in Sec.~IV, several surveys were given during the semester to gauge
the students' perceptions of the course and to provide feedback on
further improvements.

A brief pre-assessment was given on the second day of class and two
more detailed assessments were given in the last week of class, in
addition to several unofficial feedback surveys during the semester.
The assessments and surveys show that the students found the course
valuable and that many of the goals in the syllabus were met. A typical
student comment was {}``I really enjoyed the class. Not only did
I learn about what advances have been achieved (or will be soon),
but also the social implications towards using/creating technology.''

\begin{figure}
\begin{center}\includegraphics[%
  scale=0.4]{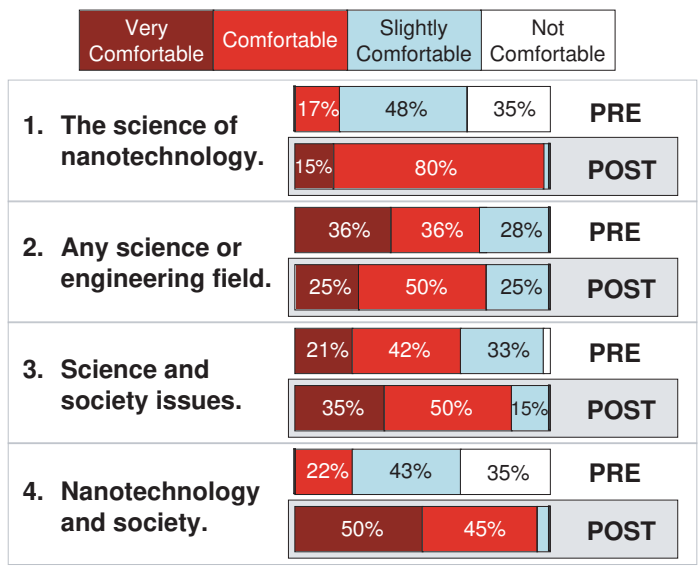}\end{center}

\caption{Pre- and post-assessment answers to the question: {}``Please rate
your comfort level with the following topics.''\label{cap:PrePost}}
\end{figure}

The pre-assessment attempted to gauge the comfort and knowledge levels
of the topics to be studied in the course as well as of nanoscale
science and engineering in general. Figure~\ref{cap:PrePost} shows
the results of the comfort level assessment before and after the bulk
of the course.%
 Of note is the general increase in comfort level for all topics and
the improvement in the area of nanotechnology and society. By the
end of the course 95\% of the class claimed to be {}``comfortable''
or {}``very comfortable'' with the subject, a tremendous improvement.
In addition, the pre-assessment asked the students to define nanotechnology
and list several nanotechnologies that they knew, as well as whether
and where they had heard the term. About a quarter of the class said
that this course was the first time they had heard the term. The others
cited news, TV, or science fiction as their source of introduction.
Initially, most students described nanotechnology as {}``tiny,''
{}``microscopic,'' or {}``advanced.'' The most common answers
were variations on {}``the study of small particles or very small
technology'' or circular definitions such as the {}``study/design/manufacturing
of products/objects at the nanoscale.'' Only one student cited $1\times10^{-9}$
meters as a benchmark. Before the course students cited {}``advanced/really-fast
computers'' as the most common example for nanotechnology, followed
by {}``medical/medicine,'' and {}``stain free pants.''

The final exams and post-assessment asked these same questions again
plus more in-depth questions about the students' knowledge of nanoscale
science and engineering. When asked to define nanotechnology, almost
all the students were able to give a working definition of nanoscale
science and engineering on par with or surpassing the definitions
found elsewhere. The students also could cite examples of new phenomena
that occur at the nanoscale including increased reactivity, quantum
confinement effects, and biological coincidences (such as the ability
of nanoparticles to cross the blood-brain barrier), as well as more
specific examples. All the students were able to give three examples
of specific nanotechnologies. Moreover, the students were able to
formulate three meaningful questions about the societal implications
of nanoscale science and engineering, a question on the pre-assessment
that was left mostly blank.

The post-assessment included additional questions to judge the impact
of the course on the students. The students were asked to summarize
the class in a sentence or two; the following comment is representative.
{}``This class gave me a good overview of the science of nanotech
and its societal implications. I now feel much better about current
trends in the field.''

To fully interpret the post-assessment results, it is useful to revisit
the students' backgrounds and motivations. Many of the students (14)
took the class to fulfill a humanities requirement with about half
also citing a general interest in nanotechnology. Out of 22 total
students, roughly two-thirds did not come from a humanities background
but instead came from the engineering and natural sciences, business,
and related fields. Out of five women and seventeen men, there were
four freshman, ten sophomores, three juniors, and five seniors. The
largest contingent from any one major was from biochemistry (4) followed
by computer science (3).

Fourteen students would take the course again even if it didn't fulfill
a requirement, although a quarter would not. Nearly all (17 yes, 3
maybes) would recommend the course to another student. All said their
knowledge of the science of nanoscale science and engineering improved
because of this course. One student commented: {}``I knew very little
about nanotechnology and I was surprised by how much there is.''
Nearly all (17) said the course made them very or extremely well prepared
to explain what nanoscale science and engineering is. For example,
one comment stated that the course {}``provides a basic, layman's
definition as well as an in-depth definition.'' Nearly all (18) considered
`nanotechnology and society a valuable field'' of intellectual pursuit,
which was somewhat surprising to us considering the newness and ambiguity
of the field when we started.

Before the course, most students were planning on pursuing a career
in science and engineering (3 were not, 2 maybe), and none were considering
one in nanotechnology. Students were largely not encouraged to change
to a more nano-related career (8 maybe, 10 no), but the course encouraged
them to be aware of opportunities and relations to nanoscale science
and engineering in their planned field (15 yes). The course did not
encourage the students to pursue a career in STS or policy (5 maybe,
16 no). Three-quarters of the class said that their perspective on
science, technology, and societal implications changed as a result
of the class. A typical student comment was that {}``Before the course,
I thought any/all technological improvements were good. Now I understand
more of the social issues of new technology.''

Most of the students thought the class was sufficiently challenging,
although a few expected more and most thought the course could not
or only might be improved significantly. About a quarter of the students
would have liked to see more science, about a quarter thought there
was too much, and 50\% thought it was a good mix. The students preferred
in-class activities, debates, town-hall meetings, and generally doing
the work themselves over traditional lectures. The research project
presentations were universally thought to be a good idea, but the
students would have preferred more specificity and direction from
the instructor.

Finally, the essay assignments provided a means to apply and test
the application of higher order analytical skills and concepts to
present day issues in nanotechnology and society. Although assessment
cannot be quantitative in this regard, we found that the students
did reasonably well (with some variation in skill level) in thinking
creatively and knowledgeably on the issues in question. Not only did
they show a growing understanding of how nanotechnology will affect
society (with past technologies as test cases), but how society can
determine the evolution and application of technology (see Table~\ref{cap:Essays}).

A rewarding message from the post-assessment and in-class surveys
was that the students overwhelmingly preferred discussion/group-oriented
classes over lecture-oriented classes. {}``Some of the more science
based aspects are taught better in lecture format. This was done for
the main part. But implications on society is better in discussion
format.'' Another good point was {}``nanotech is changing so fast,
it'd be bad to try and follow a pre-established lecture schedule.''

\section{Discussion and Reflection}

A social science course that focuses on technology creates unique
challenges and new opportunities for education. With over half the
class composed of science or engineering majors, there was a bias
against the more open-ended, subjective questions that can be posed
in science and technology studies. Many students expected a class
about nanotechnology.

Clarity is the first step in good student engagement. The philosophy
and content of the course must be clearly and repeatedly explained,
focusing on why the subject is worthwhile and what will be gained
from a significant time investment. The instructor's (CT) technical
background helped somewhat in that it gave credibility and a starting
point for a new direction of intellectual pursuit. In the end though,
personal attention --- learning the students' names, majors, career
plans, interests --- is necessary to enlist the class in learning,
especially in the context of group work, class participation, and
active learning activities. Not surprisingly, this attention requires
much effort on the instructor's part. It is also tremendously rewarding.

Teaching the course required a lot of leadership. We pushed and pulled
in different directions as the course navigated through various paces
and types of content. We bounced back and forth between STS and nanoscale
science and engineering to keep student interest and integrate concepts
and theories. Because the course was offered for the first time, extra
preparation was needed for each class. The course schedule was also
quite fluid as the order and depth of the course material was continually
calibrated to match the students' learning pace and the instructors'
growing experience.

We had thought the students would be mostly in their first year. Instead,
we attracted a much more diverse and older student body. Older students
with science and engineering majors tend to be more resistant to active
learning techniques and class participation. They are also more competent
overall, be it in writing, reading, or analytical comprehension abilities,
which can lead to boredom in mixed skill-level environments. We made
this overqualification into an opportunity. The research projects
and essay assignments provided a good way to challenge the students
while keeping everyone engaged at their ability level. The nano research
projects became continuing educational tools for both the researcher
and the rest of the class in research and communication techniques
as well as general knowledge.

So how much work did it take? For the students, a balance had to be
maintained between university requirements and their expectation and
commitment level. The class decided collectively to meet as groups
in-class but have individual homework and assignments outside of class.
For important concepts or theories in STS, the class settled into
a routine of working in groups on work sheets or quizzes provided
by the instructor, then as a class reviewing their work. The nanoscience
discussions tended to be more whole class oriented with individual
students contributing their research or perspective. After the learning
goals were set by the instructor, the class preferred to work in small
groups. The amount of work required on the students part was similar
to other courses at the university.

The instructor had more extensive duties. In addition to preparing
for a course with no standard text for the first time, the research
projects required special attention. The students learned more about
nanoscale science and engineering through the projects and applied
their newfound societal analytical toolset to explore the implications
of their nano-topic. The instructor's philosophy was to model the
progress and requirements of the project on a real-world research
group, where the students would need to meet milestones and share
their progress with the rest of the class at group meetings. The formal
class presentation was a step in this process of producing a readable
report. The implementation of this approach was good but not perfect.
Some of the students would have benefited from more hand-holding and
specification. Despite the instructor's not limitless time, the assessments
showed that the experience was found to be valuable by almost all
of the students. In summary, realistic time constraints were not a
barrier to preparing and teaching an effective and interesting course
from our perspective.

Scientists and technologists, as well as science students, consider
the societal ramifications of technology all the time. Well, at least
they should. But thinking critically about such issues in a course
involving science and technology studies, history of science, and
public policy professionals is generally a new and very worthwhile
experience. An exciting new field of study like nanotechnology can
provide the basis for learning about the issues of technological change
alongside technological developments in real-time.

\begin{acknowledgments}
We are grateful to the National Science Foundation through the Materials
Research Science and Engineering Center on Nanostructured Materials
and Interfaces (DMR-0079983) and through the Nanotechnology Undergraduate
Education (NUE) grant, \emph{An Integrated Approach to Teaching Nanotechnology
and Society} (DMR-0407075). Both programs are at the University of
Wisconsin-Madison.

We would like to thank the EPD 690 students for their help in thinking
about the complex issues surrounding nanotechnology and society. In
particular, we would like to thank Anne Bentley and Adam Creuziger
for participating in the second portion of the course and creating
syllabi and other course materials. CT would like to thank Robert
Joynt for allowing him to develop and teach this course while pursuing
his Ph.D.\ in physics. 
\end{acknowledgments}

\end{document}